%% file: main.tex
\documentclass[conference,10pt]{IEEEtran}
\IEEEoverridecommandlockouts
\usepackage{cite}
\usepackage{amsmath,amssymb,amsfonts}
\usepackage{algorithmic}
\usepackage{graphicx}
\usepackage{textcomp}
\usepackage[table]{xcolor}

\usepackage{hyperref}
\usepackage{multirow}
\usepackage{threeparttable}
\usepackage{subcaption}
\usepackage{booktabs}
\usepackage{graphicx}
\usepackage{amssymb,amsmath}
\usepackage{adjustbox}
\usepackage{enumitem}
\usepackage{bbding}
\usepackage{stfloats}
\usepackage{url}
\usepackage{cite}

\usepackage{soul}  
\newboolean{showedit}
\setboolean{showedit}{true}  
\newcommand{\maybehide}[1]{%
  \ifthenelse{\boolean{showedit}}%
    {\st{#1}}%
    {}%
}
\newcommand{\alextodo}[1]{%
  \ifthenelse{\boolean{showedit}}%
  {\textcolor{blue}{#1}} 
  {#1}
}

\def\BibTeX{{\rm B\kern-.05em{\sc i\kern-.025em b}\kern-.08em
    T\kern-.1667em\lower.7ex\hbox{E}\kern-.125emX}}
\begin{document}

\title{USAD: Universal Speech and Audio \\ Representation via Distillation
}

\author{\IEEEauthorblockN{Heng-Jui Chang, Saurabhchand Bhati, James Glass, Alexander H. Liu}
\IEEEauthorblockA{
    \textit{MIT CSAIL} \\
    Cambridge, MA, USA \\
    hengjui@mit.edu
}


}

\maketitle

\begin{abstract}
Self-supervised learning~(SSL) has revolutionized audio representations, yet models often remain domain-specific, focusing on either speech or non-speech tasks.
In this work, we present Universal Speech and Audio Distillation~(USAD), a unified approach to audio representation learning that integrates diverse audio types—speech, sound, and music—into a single model.
USAD employs efficient layer-to-layer distillation from domain-specific SSL models to train a student on a comprehensive audio dataset.
USAD offers competitive performance across various benchmarks and datasets, including frame and instance-level speech processing tasks, audio tagging, and sound classification, achieving near state-of-the-art results with a single encoder on SUPERB and HEAR benchmarks.\footnote{Models: \url{https://huggingface.co/MIT-SLS/USAD-Base}}
\end{abstract}



\input{section/intro}
\input{section/related}

\input{section/method}

\input{section/exp}

\input{section/conclusion}

\bibliographystyle{IEEEtran}
\bibliography{refs,refs_audio}

\end{document}

%% file: section/intro.tex
\section{Introduction}
\label{sec:intro}

In recent years, self-supervised learning~(SSL) methods—learning frameworks that utilize unlabeled data without explicit supervision—have significantly advanced representation learning for audio processing.
Speech SSL models like wav2vec 2.0~\cite{baevski2020wav2vec2}, HuBERT~\cite{hsu2021hubert}, and WavLM~\cite{chen2022wavlm} have become the foundation of many applications like automatic speech recognition~(ASR), speaker identification, and phoneme classification.
In parallel, SSL approaches developed for audio event classification and music understanding, such as SSAST~\cite{gong2022ssast}, BEATs~\cite{chen2022beats}, and MERT~\cite{li2023mert}, have successfully been shown to be effective in non-speech tasks.

In practice, the use of audio representation has extended beyond simple downstream tasks.
These representations have become the proxy for modern multimodal systems to understand the world and to interact with humans through audio.
For example, many audio-enabled large language models~(LLM) are trained to understand audio via representations~\cite{gong2023ltu,tang2024salmonn,chu2024qwen2audio,ghosh2024gama}, where researchers have found the quality of the representation to be critical to audio LLM performance~\cite{zhang2023usm}.
Furthermore, audio representation learning has been applied to new encoding techniques that empower complex systems to understand and generate speech~\cite{zhang2023speechtokenizer,borsos2023soundstorm,defossez2024moshi,chang2025dcspin}.

Despite these advancements, open challenges remain in audio representation learning.
In this work, we are particularly interested in the problem of learning representations of general audio—the union of speech, sound, and music.
While these different types of audio are essentially signals with various patterns, they have been treated as separate domains in the audio processing literature (see Section \ref{sec:related}), with specialized models tailored specifically to each.
On the road to artificial general intelligence, audio is an inevitable barrier that needs to be solved regardless of the context.
Having disjoint representations for different types of audio increases the complexity of solutions towards the ultimate goal~\cite{gong2023ltuas,chu2024qwen2audio,wang2025usam}.

To address this problem, we propose Universal Speech and Audio Distillation (USAD), a unified model for speech and audio representation.
USAD is trained via layer-to-layer distillation~\cite{chang2024colld} from pre-trained teacher models in speech and audio, using a mixture of multi-domain audio data.
Our results show that USAD is competitive across various benchmarks in speech, sound, and music processing, often approaching state-of-the-art performance despite being a single, general-purpose audio encoder.
Additionally, we conduct ablation studies on key method elements, including teacher model choice, student model size, distillation data, and computational budget.
To summarize, our contribution can be highlighted as follows:
\begin{itemize}
    \item We propose USAD, a unified audio representation model that jointly learns from speech, sound, and music.
    \item USAD distills domain-specific experts to an encoder.
    \item Results show USAD learns a unified embedding space across domains with a reasonable computational budget.
    \item Notably, USAD encodes audio into representations that closely match the performance of oracle teacher models and state-of-the-art expert models in diverse audio tasks, regardless of input domain.
\end{itemize}

%% file: section/related.tex
\section{Related Works}
\label{sec:related}

\input{figure/framework}

\noindent \textbf{\textit{Speech Representation Learning.}}
To utilize large-scale unlabeled speech datasets, well-known SSL models like wav2vec 2.0~\cite{baevski2020wav2vec2} and HuBERT~\cite{hsu2021hubert} rely on predicting pseudo labels (derived from contrastive learning or offline clustering) of partially masked speech to train the representation encoder.
These methods have shown that learning through pseudo labels significantly improves downstream performance on speech tasks compared to a fully supervised method without SSL.
WavLM~\cite{chen2022wavlm} further improves the technique by introducing a denoising task to SSL training, thus producing richer representations useful across a broader range of speech applications such as ASR, keyword spotting, and speaker recognition.

\noindent \textbf{\textit{Sound and Music Representation Learning.}}
Similar to prior works in speech, sound/music SSL frameworks are designed to learn semantically meaningful representations from large-scale, unlabeled datasets.
Reminiscent of speech SSL methods, prior works in this area found the combination of pseudo labeling and masked prediction effective~\cite{gong2022ssast,baade2022mae,niizumi2022masked,huang2022masked}.
Not surprisingly, the resulting representation enhances audio processing tasks like sound and music classification.
The inspiration for this work stemmed from the similarities shared by speech and non-speech representation learning frameworks.

\noindent \textbf{\textit{Self-supervised Learning with Distillation.}}
Distillation has been studied extensively in audio SSL and can be divided into self-distillation~(SD) and knowledge distillation~(KD), where the difference lies in the teacher model.
For SD~\cite{baevski2022data2vec,baevski2023data2vec2,liu2023dinosr,chen2024eat,alex2025sslam,niizumi2021byol,li2024atst,niizumi2023masked}, teacher and student models have the same size and are randomly initialized.
SD relies on providing different views to the two models~(e.g., masked and unmasked) and designing an update policy for the teacher model~(e.g., moving average).
Meanwhile, KD relies on pre-trained teacher models as the source of knowledge.
Methods like DistilHuBERT~\cite{chang2022distilhubert}, DPHuBERT~\cite{peng2023dphubert}, DASS~\cite{bhati2024dass}, and CoLLD~\cite{chang2024colld} utilize this technique to compress large models.
DistilHuBERT focuses explicitly on reducing model complexity by predicting hidden representations from a teacher model layer by layer, achieving comparable accuracy at significantly lower computational cost.
Multi-teacher ensemble distillation has also been explored previously, but is limited to speech~\cite{huang2023ensemble}.
Another work uses decoupled KD and task arithmetic to learn a unified compressed speech and music encoder~\cite{ritter2025distilling}.

While USAD falls in the category of KD, USAD differs from existing works: (1) USAD simultaneously distills from two teacher models from different domains; (2) USAD aims at generalization rather than compression of the student model; (3) existing approaches remain confined to speech tasks, without considering cross-domain audio representation learning.

%% file: figure/framework.tex
\begin{figure*}[t]
    \centering
    \includegraphics[width=0.87\linewidth]{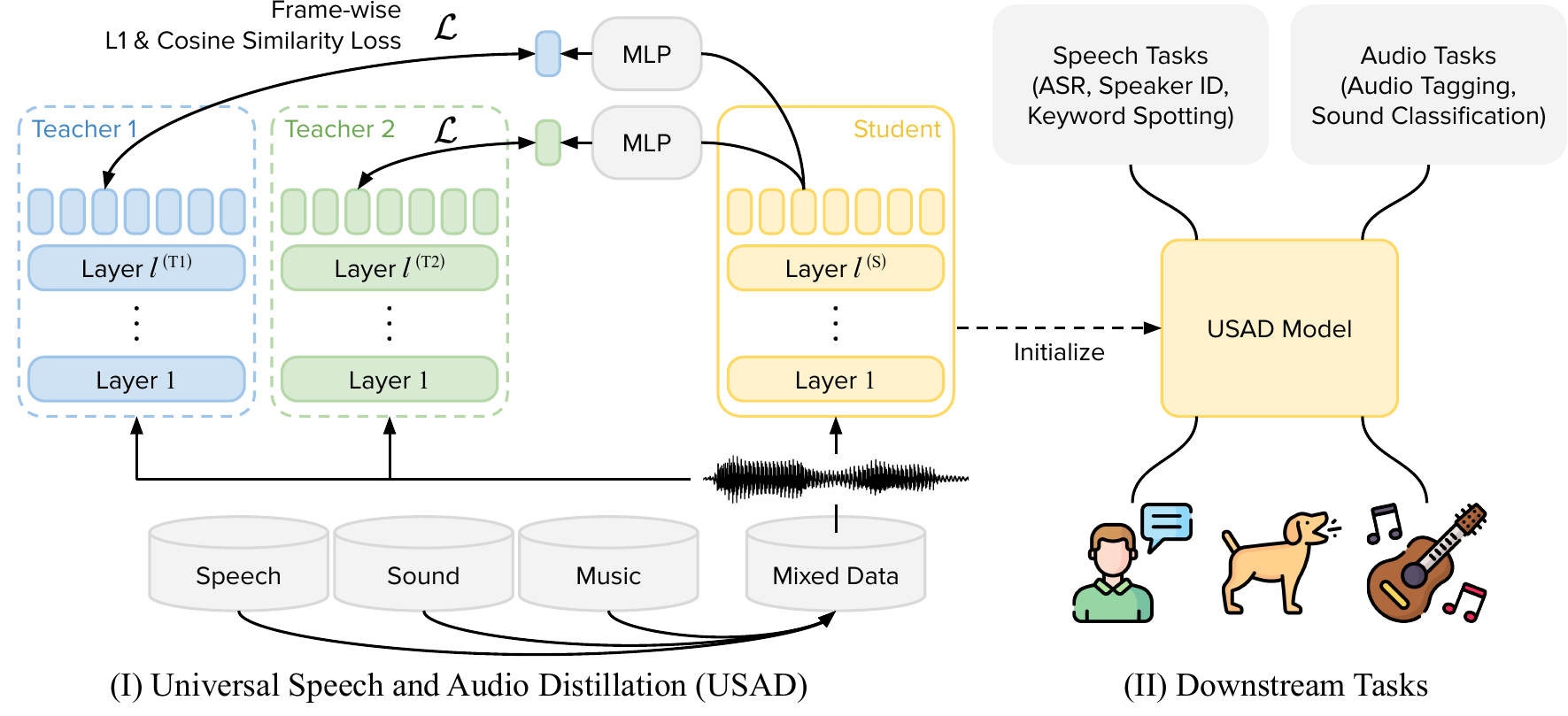}
    \vspace{-7pt}
    \caption{
        The proposed USAD.
        In stage~(I), a student model distills knowledge from two domain-specific teachers through sparse layer-to-layer distillation with mixed audio data.
        In stage~(II), the USAD model can be used in various speech and audio tasks.
    }
    \label{fig:framework}
    \vspace{-9pt}
\end{figure*}

%% file: section/method.tex
\section{Methods}
\label{sec:method}

\subsection{Unified Speech and Audio Distillation}
\label{sec:method-usad}

This paper aims to develop a unified encoder to extract representations across speech, sound, and music domains.
Motivated by the success of single-domain SSL and knowledge distillation, we propose Unified Speech and Audio Distillation~(USAD) to distill domain-specific SSL pre-trained models into a shared encoder.
The framework is depicted in Fig.~\ref{fig:framework}.
First, we select two pre-trained SSL models (T1 and T2) that were trained on different domains.
A randomly initialized student model and multi-layer perceptron~(MLP) prediction heads are used.
During training, all models receive the same mixed dataset inputs, comprising speech, sound, and music.
Specific student layer~($l^{(\text{S})}$) outputs are fed into two MLP heads to predict corresponding hidden layer~($l^{(\text{T1})}$ and $l^{(\text{T2})}$) representations from the teacher models.
As all models use transformer encoders~\cite{vaswani2017attention}, the student layers predict the teacher's feed-forward network~(FFN) features, which is effective as demonstrated previously~\cite{baevski2022data2vec,chang2024colld}.
We use the frame-wise reconstruction loss from DistilHuBERT~\cite{chang2022distilhubert}, detailed further in the following section.

\subsection{Sparse Layer-to-layer Distillation}
\label{sec:method-l2l}

Unlike most distillation works, USAD adopts two teacher models to learn general representations across multiple audio domains.
Designing an efficient distillation method to train the student model is thereby a key factor to make USAD useful in practice.
We propose sparse layer-to-layer~(L2L) distillation that improves the efficiency of the existing~(dense) L2L distillation methods with a simplified objective function.

L2L distillation, proposed by CoLLD~\cite{chang2024colld}, is originally designed to train the student network to predict the output of \textit{each and every} layer of the teacher model.
While L2L distillation demonstrates strong results, the exhaustive distillation scheme inevitably leads to increased computational cost.
To mitigate this issue, we introduce sparse L2L distillation that distills
only $K$ layers, exploiting the fact that similarity between consecutive transformer layers is expected to be high~\cite{pasad2021layer}.
For a 12-layer student with $K=4$, distillation occurs at $l^{(\text{S})} \in \{3,6,9,12\}$, reducing loss compute by approximately 75\%.
Formally, given student and teacher model depths $L^{(\text{S})}$, $L^{(\text{T1})}$, and $L^{(\text{T2})}$, the $k$-th student layer $l^{(\text{S})}_k = \left\lfloor \frac{kL^{(\text{S})}}{K} \right\rfloor$ is matched to teacher layers
$l^{(\text{T1})}_k = \left\lfloor \frac{kL^{(\text{T1})}}{K} \right\rfloor$ and
$l^{(\text{T2})}_k = \left\lfloor \frac{kL^{(\text{T2})}}{K} \right\rfloor$, where $k = 1, \dots, K$.
The student's output at $l^{(\text{S})}_k$ passes through two separate MLP heads, which predict the corresponding FFN features of each teacher.

In addition to sparsity introduced in the learning target, we improve the training objective of L2L in the same spirit.
The original design of CoLLD utilizes contrastive loss that requires both positive and negative samples on a per-frame basis, making loss computation expensive.
To reduce the cost, we eliminate the need for negative samples by adopting L1-cosine similarity from DistilHuBERT~\cite{chang2022distilhubert}.
For predicting teacher T1's FFN features from student layer $l^{(\text{S})}_k$, we define the student's MLP output:
$\tilde{\boldsymbol{z}}^{(\text{T1})}_{k,t} = f_{k}^{(\text{T1})} \left(\boldsymbol{h}^{(\text{S})}_{k,t} \right) \in \mathbb{R}^D$.
The training objective minimizes L1 distance and maximizes cosine similarity between predictions and target features:
\vspace{-4pt}
\begin{equation*}
    \mathcal{L}_{k,t}^{(\text{T1})} =  \frac{1}{D} \left\| \tilde{\boldsymbol{z}}^{(\text{T1})}_{k,t} - \boldsymbol{z}^{(\text{T1})}_{k,t} \right\|_1 - \log\sigma \left[ \cos \left( \tilde{\boldsymbol{z}}^{(\text{T1})}_{k,t} , \boldsymbol{z}^{(\text{T1})}_{k,t} \right) \right],
    \vspace{-2pt}
\end{equation*}
where $\sigma$ is the sigmoid function.
We aggregate losses across layers and frames:
\vspace{-7pt}
\begin{equation*}
    \mathcal{L} = \frac{1}{2KT} \sum_{k=1}^K \sum_{t=1}^T \mathcal{L}_{k,t}^{(\text{T1})} + \mathcal{L}_{k,t}^{(\text{T2})},
    \vspace{-3pt}
\end{equation*}
where $T$ is the number of audio frames.
This setup significantly improves computational efficiency compared to CoLLD, preserving comparable performance.

\subsection{Audio Feature Extraction: Frame vs. Patch}
\label{sec:method-feature-extract}

A critical distinction between speech and audio SSL is the feature extraction strategy.
Phonetic durations are typically on the order of 100 ms, necessitating a temporal resolution of at least 10Hz, and making speech processing tasks require fine-grained features.
Thus, speech models utilize frame-wise features—audio segments downsampled along the time axis~(e.g., 50Hz in models like WavLM~\cite{chen2022wavlm} and Whisper~\cite{radford2022whisper}).

In contrast, general audio signals exhibit complex temporal and frequency patterns, primarily targeting instance-wise classification tasks.
Therefore, audio SSL approaches often use lower temporal resolution encoders or patch-based features~\cite{li2024atst,gong2022ssast,chen2022beats,chen2024eat,alex2025sslam}.
Typically, these methods begin with a 128-dimensional Mel-spectrogram with a stride of 10 ms, cropped into non-overlapping 16$\times$16 patches, and flattened into a sequence of 256-dimensional vectors.
Although the patch sequence has a 50Hz framerate, the actual temporal resolution is 6.25Hz, smoothing out local details and blurring speech-specific information.

While patch-based features preserve frequency detail, this approach degrades speech task performance~(see Section \ref{sec:exp}).
Moreover, despite having the same framerate, frame- and patch-based teacher features are misaligned, complicating the student's joint representation learning.
Thus, we propose distilling from frame-based audio teachers to optimize performance across both speech and audio domains.

%% file: section/exp.tex
\section{Experiments}
\label{sec:exp}

\input{table/dataset}
\input{table/usad_hparam}

\subsection{Setup}
\label{sec:exp-setup}

\noindent\textit{\textbf{Self-supervised Models.}}
For speech SSL models, we include DPWavLM~\cite{peng2023dphubert}, WavLM Base+~\cite{chen2022wavlm}, and data2vec 2.0 Speech~\cite{baevski2023data2vec2}, where DPWavLM is a compressed WavLM Base+ model.
For audio SSL models, we consider SSAST~\cite{gong2022ssast}, BEATs~\cite{chen2022beats}, EAT~\cite{chen2024eat}, SSLAM~\cite{alex2025sslam}, and ATST~\cite{li2024atst}, where SSAST and ATST both have a frame-based version, but ATST has a framerate of 25Hz.
Additionally, we pre-train two data2vec 2.0 models of the same architecture as the speech version, using both audio and mixed data for comparison.

\noindent\textit{\textbf{Data.}}
To maximize domain coverage, we combine multiple datasets containing English speech, sound, and music, as shown in Table~\ref{tab:dataset}.
First, we segment sound and music into 10-second clips, discarding segments shorter than two seconds or longer than 30 seconds for training efficiency, and remove silence clips.
All clips are resampled to 16kHz.
To balance speech and non-speech data, audio and music datasets are upsampled by a factor of two.
This balanced dataset is named \textbf{Mix126k-B}, where ``126k'' indicates the pre-upsampled duration in hours.
Additionally, we create a smaller dataset for ablation studies by downsampling LibriVox~(LV) to match AudioSet~(AS), denoted as \textbf{LV-AS}.
Note that LV is LibriSpeech~\cite{Panayotov2015libri}, Libri-Light~\cite{kahn2020libri}, and Multilingual LibriSpeech~\cite{pratap2020mls} combined with no overlapping utterances.

\noindent\textit{\textbf{USAD Model and Training.}}
USAD is implemented with fairseq~\cite{ott2019fairseq} and trained on four NVIDIA A6000 GPUs.
First, the input waveform is transformed into a 128-dimensional Mel-spectrogram~(25 ms window, 10 ms stride) and normalized following SSAST~\cite{gong2022ssast}.
If any of the teacher models use frame-wise features, we add a convolutional feature extractor with a stride of two.
For patch-wise features, a patch embedding module similar to those used in standard audio SSL approaches is employed.
The features are then normalized~\cite{ba2016layer}.
If the two teachers have different types of features, we sum the extracted features since they share the same framerate.
The features are fed to a 5-layer convolutional positional encoding as in data2vec, then input to a transformer encoder with relative positional encoding~\cite{shaw2018relative-pos}.
Each prediction head consists of two fully-connected layers with a ReLU activation in between.
We use a linear learning rate scheduler with warmup.
Detailed hyperparameters for USAD are listed in Table~\ref{tab:usad-hparam}.
When ATST is one of the teachers, we apply mean pooling with a kernel size and stride of two before calculating the loss, as ATST has a lower framerate.

\input{table/speech_audio_eval}

\noindent\textit{\textbf{Downstream Tasks.}}
We evaluate USAD on SUPERB~\cite{yang2021superb,tsai-etal-2022-superb,yang2024large} and HEAR~\cite{turian2022hear}.
These benchmarks freeze the model to be evaluated, extract the hidden layer representations, and train a lightweight prediction head for each task.
Additionally, we fine-tune the encoders for audio tagging~(AS-20K)~\cite{gemmeke2017audioset} and sound classification~(ESC-50)~\cite{piczak2015esc50}, where AS-20K is a subset of AudioSet containing 20k recordings with balanced labels over all 527 possible sound classes and 5-fold cross-validation is employed for ESC-50.

To offer an overall performance metric across multiple tasks, we compute the SUPERB Score~\cite{feng2022superb} for each model $f$:
\vspace{-4pt}
\begin{equation*}
    \text{SUPERB\_Score}(f) = \frac{1000}{|\mathcal{T}|} \sum_{\tau \in \mathcal{T}} \frac{s_{\tau}(f) - s_{\tau}(\text{baseline})}{s_{\tau}(\text{SOTA}) - s_{\tau}(\text{baseline})},
\end{equation*}
where $\mathcal{T}$ is the set of tasks and $s_{\tau}(\cdot)$ is a model's measured performance of task $\tau$, e.g., $s_{\tau} =$ WER and $\tau =$ ASR.
The baselines are fbank features for SUPERB and randomly initialized SSAST for audio tasks~\cite{gong2022ssast}.

\subsection{Downstream Speech Evaluation}
\label{sec:exp-speech-eval}

We first evaluate models on SUPERB by adopting three frame-level tasks: phoneme recognition~(PR), automatic speech recognition~(ASR), and speaker diarization~(SD); and four instance-level tasks: keyword spotting~(KS), intent classification~(IC), speaker identification~(SID), and emotion recognition~(ER).
As shown in Table~\ref{tab:speech-audio-eval}, the USAD Small model surpasses the similar-size DPWavLM on most SUPERB tasks.
The performance gap relative to the WavLM teacher narrows as the USAD model size increases, surpassing the SUPERB Score in instance-level speech tasks for USAD Large.
USAD outperforms all audio SSL models listed in Table~\ref{tab:speech-audio-eval}, especially in frame-level tasks, highlighting the advantage of distillation from a speech SSL teacher.
Therefore, USAD demonstrates strong capabilities in speech processing tasks.

\subsection{Downstream Audio Evaluation}
\label{sec:exp-audio-eval}

Fine-tuning for sound event detection and classification has been a standard evaluation for audio SSL models, so we follow EAT to fine-tune USAD models~\cite{chen2024eat}.\footnote{\url{https://github.com/cwx-worst-one/EAT}}
Following SSAST~\cite{gong2022ssast}, we average the weights from 10k, 15k, $\dots$, 40k updates checkpoints.
As shown in Table~\ref{tab:speech-audio-eval}, USAD models outperform all speech and some audio SSL models in audio tasks, despite the gap between USAD and the ATST Frame teacher remaining mainly because USAD is designed to perform well on both speech and general audio tasks.
Patch-based approaches dominate sound event classification, but USAD offers steady improvements by scaling up model capacity, surpassing the ATST Frame teacher's Audio SUPERB Score with the Large model.
Moreover, because speech and audio SSL require different masking strategies, data2vec 2.0 Mix is more challenging to learn general domain representations, resulting in inferior performance compared with USAD.
Combined with the findings in Section~\ref{sec:exp-speech-eval} and the average SUPERB Score, USAD achieves the best overall results, demonstrating the success of the proposed methods.

\subsection{Joint Speech and Audio Evaluation}
\label{sec:exp-joint-eval}

\input{table/hear_benchmark}

This section compares USAD with the teacher models on HEAR, a comprehensive benchmark that encompasses 19 tasks covering speech, sound, and music.
In Table~\ref{tab:hear2021}, the WavLM teacher performs poorly because most tasks relate to sound and music, showing the necessity of building a joint speech and audio representation model.
Next, we set a topline by concatenating the representations of the two teacher models, denoted as ``WavLM + ATST.''
Compared to the topline, USAD shows superior performance in speech tasks like CREMA-D~(speech emotion classification) and VoxLingua107~(language identification).
In contrast, the degradation in sound and music tasks suggests that USAD models may focus more on distilling information from the WavLM teacher.
The average performance is improved when USAD is scaled to larger sizes, surpassing the topline and closing the gap with state-of-the-art results.
To summarize, the HEAR benchmark offers a comprehensive indicator of the effectiveness of USAD, while also highlighting potential advancements.

\subsection{Teacher Model Selection}
\label{sec:exp-teacher-select}

\input{table/ablation_teacher}

This section discusses the effect of the teacher models for USAD training.
All models follow the USAD Small Ablation setup in Table~\ref{tab:usad-hparam}, trained with the LV-AS dataset mentioned in Section~\ref{sec:exp-setup}.
In Table~\ref{tab:ablation-teacher}, USAD with one teacher performs well in either the speech or audio domain, implying the necessity of distilling from multiple SSL teachers specializing in different domains.
Next, we explore combinations of speech and audio teachers.
Compared with frame-based audio teachers, patch-based models degrade PER while offering similar AS-20K results, corroborating our hypothesis that frame-based feature extraction is more suitable to jointly learn speech and audio representations because the learning targets from the two teachers are aligned~(Section~\ref{sec:method-feature-extract}).
A shared phenomenon is the apparent trade-off between PR and AS-20K results, indicating that the choice of teacher models remains a crucial factor in USAD training.

\subsection{Data Distribution}
\label{sec:exp-data-distribution}

\input{figure/utt_ratio_per_map}

We inspect the effect of USAD training data distribution by training 11 USAD models with different combinations of the datasets listed in Table~\ref{tab:dataset}.
As shown in Fig.~\ref{fig:utt-ratio-per-map}, the ratio between speech and non-speech recordings strongly correlates with downstream performance.
That is, with more speech clips present in the training data, USAD's capability in modeling speech increases~(Fig.~\ref{fig:utt-ratio-per}) but degrades for non-speech tasks~(Fig.~\ref{fig:utt-ratio-map}).
Nonetheless, the drop in phoneme recognition performance when reducing speech data is less significant than the improvements on AS-20K, suggesting that training USAD with a balanced dataset of speech and non-speech offers more generalized representations across various domains.
Consequently, we propose upsampling the sound and music data in the USAD training data to match the speech data size~(Mix126k-B in Table~\ref{tab:dataset}).

\subsection{Distillation Strategies}
\label{sec:exp-distill-strategy}

\input{table/distill_strategy}

This section discusses the effect of distillation strategies in USAD.
As shown in Table~\ref{tab:distill-strategy}, the InfoNCE loss from CoLLD yields worse results.
Under identical conditions, we found replacing InfoNCE with L1-cosine similarity increases training speed by 25\%, suggesting the L1-cosine similarity loss is more suitable for USAD.
With different distillation objectives, we found that mask-prediction helps PR performance but degrades AS-20K.
We suspect that the difference in masking strategies for speech and non-speech SSL is the cause of this issue.
Moreover, we inspect the effect of the layer-to-layer distillation technique by varying $K$, i.e., the number of layers to which the distillation loss is applied~(Section~\ref{sec:method-l2l}).
The results show that distilling from more hidden layers degrades downstream performance~($K=$ 12 and 6), implying that some teacher layer representations are unhelpful for the student to learn from.
Meanwhile, $K=$ 3 and 4 behave similarly, indicating that a potential improvement may be to find the optimal combination of teacher layers to distill from.

\subsection{Training Efficiency}
\label{sec:exp-train-efficiency}

\input{figure/flops_per_map}

To demonstrate the training efficiency of USAD compared to SSL approaches, we plot the training FLOPS against downstream performance.
In Fig.~\ref{fig:flops-per-map}, a distinct capacity–efficiency trade-off is shown: in phoneme recognition, all USAD variants converge at a similar rate, and extra parameters reduce the final PER, implying that the Small model already captures local acoustic cues.
In contrast, AS-20K performance scales almost linearly with model size because the task depends on long-range, semantically rich information and benefits from larger receptive fields and extra representational bandwidth, while the larger model capacity also regularizes the weak, noisy labels.
Furthermore, USAD Small surpasses data2vec 2.0 Mix with under one EFLOPS, whereas USAD Large attains 36–37\% mAP using an order-of-magnitude less compute than ATST Frame.
Overall, the results demonstrate the training efficiency of the proposed USAD and suggest that further improvements can be achieved by scaling the model size.

%% file: table/dataset.tex
\begin{table}[t]
    \centering
    \caption{
        Preprocessed dataset information.
        The dataset sizes might differ from the original ones due to preprocessing.
    }
    \label{tab:dataset}
    \begin{adjustbox}{max width=\linewidth}
    \begin{tabular}{@{~}l@{~~}r@{~~}r@{~~}r@{~}}
        \toprule
        Dataset & Clips & \shortstack{Duration\\(hours)} & \shortstack{Proportion of\\ Mix126k-B} \\
        \midrule
        \textbf{Speech (English)} \\
        ~~LibriVox~\cite{Panayotov2015libri,kahn2020libri,pratap2020mls} & 13,974,853 & 56,167.4 & 27.5\% \\
        ~~VoxPopuli~\cite{wang2021voxpopuli} & 3,051,826 & 24,084.4 & 6.0\% \\
        ~~GigaSpeech~\cite{chen2021gigaspeech} & 4,631,558 & 6,973.2 & 9.1\% \\
        ~~Common Voice 17~\cite{ardila2020commonvoice} & 1,117,555 & 1,765.3 & 2.2\% \\
        ~~Fisher~\cite{cieri2004fisher} & 1,168,612 & 1,708.5 & 2.3\% \\
        ~~VoxLingua107~\cite{valk2021voxlingua107} & 15,860 & 48.8 & 0.03\% \\
        \midrule
        \textbf{Sound} \\
        ~~AudioSet~\cite{gemmeke2017audioset} & 1,932,298 & 5,320.6 & 7.6\% \\
        ~~SoundNet~\cite{aytar2016soundnet} & 9,765,588 & 25,553.8 & 38.4\% \\
        ~~LAION-Audio-630k~\cite{wu2023laion-audio} & 1,433,319 & 3,535.7 & 5.6\% \\
        \midrule
        \textbf{Music} \\
        ~~Music4All~\cite{santana2020music4all} & 327,807 & 910.6 & 1.3\% \\
        \midrule
        \textbf{Total} \\
        ~~Speech & 23,960,264 & 90,747.6 & 47.1\% \\
        ~~Sound + Music & 13,459,012 & 35,320.7 & 52.9\% \\
        ~~Mix126k-B & \multirow{2}{*}{50,878,288} & \multirow{2}{*}{161,388.9} & \multirow{2}{*}{100.0\%} \\
        ~~$=$ Speech + (Sound + Music)$\times$2 & \\
        \bottomrule
    \end{tabular}
    \end{adjustbox}
    \vspace{-2pt}
\end{table}

%% file: table/usad_hparam.tex
\begin{table}[t]
    \centering
    \caption{
        Hyperparameters of USAD.
    }
    \label{tab:usad-hparam}
    \begin{tabular}{@{~}lcccc@{~}}
        \toprule
        Hyperparameter & \shortstack{Small \\ Ablation} & Small & Base & Large \\
        \midrule
        \textbf{Model} \\
        ~~Parameters & 24M & 24M & 94M & 330M \\
        ~~Hidden Dimension & 384 & 384 & 768 & 1024 \\
        ~~Layers & 12 & 12 & 12 & 24 \\
        \midrule
        \textbf{Training} \\
        ~~Learning Rate & 5e-4 & 8e-4 & 1e-3 & 1.5e-3 \\
        ~~LR Warmup & 4k & 8k & 8k & 8k \\
        ~~Updates & 150k & 400k & 400k & 400k \\
        ~~Batch Size (seconds) & 200 & 800 & 800 & 800 \\
        ~~GPUs & 1 & 4 & 4 & 4 \\
        \bottomrule
    \end{tabular}
    \vspace{-8pt}
\end{table}

%% file: table/speech_audio_eval.tex
\begin{table*}[t]
    \centering
    \caption{
        Results on SUPERB, AS-20K, and ESC-50.
        KS1 refers to Speech Commands v1 with 10 types of keywords, a silence class, and an unknown class~\cite{warden2018speech-commands}, while KS2 includes all 35 classes in Speech Commands v2.
        LibriSpeech and AudioSet are denoted as LS and AS, respectively.
        The best results in \textbf{bold} and the second and third best results are \underline{underlined}.
    }
    \label{tab:speech-audio-eval}
    \begin{adjustbox}{max width=\linewidth}
    \begin{threeparttable}
    \begin{tabular}{@{}l@{~}c@{~}c@{~}c@{~}c@{~}c@{~}c@{~}c@{~}c@{~}c@{~}c@{~}c@{~}c@{~}c@{~}c@{~}c@{~}c@{~}c@{~}c@{~}c@{~}c@{~}c@{}}
        \toprule
        & & & & & \multicolumn{3}{c}{\shortstack{Frame-level \\ Speech Tasks}} & & \multicolumn{5}{c}{\shortstack{Instance-level \\ Speech Tasks}} & & \multicolumn{2}{c}{\shortstack{Instance-level \\ Audio Tasks$^{\spadesuit}$}} & & \multicolumn{4}{c}{SUPERB Score$\uparrow$} \\
        \cmidrule{6-8}
        \cmidrule{10-14}
        \cmidrule{16-17}
        \cmidrule{19-22}
        & & & Feature & Frame & PR & ASR & SD & & KS1 & KS2$^{\spadesuit}$ & IC & SID & ER & & AS-20K & ESC-50 & & \multicolumn{2}{c}{Speech} & \multirow{2}{*}{Audio} & \multirow{2}{*}{Avg} \\
        Model & Params & Data & Type & Rate & PER$\downarrow$ & WER$\downarrow$ & DER$\downarrow$ & & Acc$\uparrow$ & Acc$\uparrow$ & Acc$\uparrow$ & Acc$\uparrow$ & Acc$\uparrow$ & & mAP$\uparrow$ & Acc$\uparrow$ & & Frame & Instance \\
        \midrule
        \multicolumn{4}{@{}l}{\textbf{Speech SSL}} \\
        ~~DPWavLM~\cite{peng2023dphubert} & 24M & LS & frame & 50Hz & 8.2 & 10.2 & 5.5 & & 96.3 & 97.7 & \underline{98.6} & 82.1 & 65.2 & & 23.5 & 78.8 & & 751.3 & 868.7 & $-$499.8 & 373.4  \\
        ~~WavLM Base+~\cite{chen2022wavlm} & 95M & Mix94k & frame & 50Hz & \underline{3.9}& \underline{5.6} & \textbf{3.5} & & \underline{97.4} & \underline{98.4} & \textbf{99.0} & \underline{89.4} & \textbf{68.7} & & 24.0 & 77.0 & & \textbf{946.1} & \underline{945.9} & $-$611.0 & 427.0 \\
        ~~data2vec 2.0 Speech~\cite{baevski2023data2vec2} & 94M & LS & frame & 50Hz & \textbf{3.7} & \textbf{4.9} & 6.5 & & 96.8 & 98.2 & \underline{98.4} & 80.8 & 64.1 & & 26.0 & 73.7 & & 813.6 & 870.8 & $-$791.5 & 297.6  \\
        \midrule
        \multicolumn{4}{@{}l}{\textbf{Audio SSL}} \\
        ~~SSAST Patch~\cite{gong2022ssast} & 89M & AS + LS & patch & 50Hz & 83.9 & 97.2 & 12.9 & & 96.0 & 98.0 & 25.4 & 64.2 & 59.6 & & 31.0 & 88.8 & & $-$1390.5 & 613.8 & 310.3 & $-$155.4 \\
        ~~SSAST Frame~\cite{gong2022ssast} & 89M & AS + LS & frame & 50Hz & 46.9 & 22.8 & 6.8 & & 96.7 & 98.1 & 49.3 & 80.8 & 60.5 & & 29.2 & 85.9 & & 312.7 & 725.3 & 82.5 & 373.5 \\
        ~~BEATs iter3~\cite{chen2022beats} & 90M & AS & patch & 50Hz & 36.4 & 25.9 & 5.2 & & \underline{97.7} & 98.3 & 53.4 & 57.1 & 64.5 & & 38.3 & \underline{95.6} & & 386.1 & 717.2 & \underline{903.5} & 669.0 \\
        ~~EAT~\cite{chen2024eat} & 88M & AS & patch & 50Hz & 55.0 & 25.9 & 4.7 & & 92.8 & 98.3 & 53.7 & 45.0 & 62.5 & & \underline{40.2} & \textbf{95.9} & & 331.5 & 650.9 & \underline{959.9} & 647.4 \\
        ~~SSLAM~\cite{alex2025sslam} & 88M & AS & patch & 50Hz & 56.4 & 27.8 & 4.6 & & \textbf{98.8} & 98.1 & 51.6 & 42.6 & 62.6 & & \textbf{40.9} & \textbf{95.9} & & 294.8 & 655.3 & \textbf{993.3} & 647.8 \\
        ~~ATST Frame~\cite{li2024atst} & 86M & AS & frame & 25Hz & 20.4 & 18.8 & 4.7 & & 95.1 & 98.1 & 85.4 & 69.8 & 64.4 & & \underline{39.0} & 91.1 & & 597.6 & 806.9 & 616.9 & 673.8 \\
        ~~data2vec 2.0 Audio & 94M & AS & frame & 50Hz & 22.8 & 15.1 & 4.5 & & 95.1 & 98.0 & 85.3 & 61.1 & 64.1 & & 35.1 & 91.0 & & 658.8 & 779.1 & 536.4 & 658.1 \\
        \midrule
        \multicolumn{4}{@{}l}{\textbf{Speech \& Audio Mixed SSL}} \\
        ~~data2vec 2.0 Mix & 94M & Mix126k-B & frame & 50Hz & 6.3 & 8.4 & 4.8 & & 97.1 & 98.2 & 98.2 & 72.1 & 66.8 & & 33.1 & 87.8 & & 824.9 & 872.0 & 284.7 & 660.5 \\
        \midrule
        \multicolumn{6}{@{}l@{~}}{\textbf{Proposed (Teachers: WavLM Base+ \& ATST Frame)}} \\
        ~~USAD Small & 24M & Mix126k-B & frame & 50Hz & 7.8 & 9.5 & 4.9 & & 96.8 & \underline{98.4} & 95.5 & 73.5 & 66.3 & & 34.5 & 89.3 & & 796.2 & 871.7 & 411.0 & \underline{692.9}  \\
        ~~USAD Base & 94M & Mix126k-B & frame & 50Hz & 5.1 & 7.7 & \underline{4.2} & & 97.1 & \textbf{98.5} & 98.3 & \underline{88.6} & \underline{68.0} & & 35.7 & 91.1 & & \underline{868.9} & \underline{938.0} & 554.2 & \underline{787.0}  \\
        ~~USAD Large & 330M & Mix126k-B & frame & 50Hz & \underline{4.0} & \underline{6.5} & \underline{3.9} & & 97.1 & \textbf{98.5} & \underline{98.4} & \textbf{91.2} & \underline{68.4} & & 37.4 & \underline{92.7} & & \underline{913.4} & \textbf{948.3} & 693.3 & \textbf{851.7} \\
        \bottomrule
    \end{tabular}
    \begin{tablenotes}[flushleft]
        \item
            $^{\spadesuit}$ The SSL models are fine-tuned with downstream task data.
    \end{tablenotes}
    \end{threeparttable}
    \end{adjustbox}
    \vspace{-8pt}
\end{table*}

%% file: table/hear_benchmark.tex
\begin{table*}[t]
    \centering
    \caption{
        HEAR 2021 Benchmark~\cite{turian2022hear}.
        All metrics are higher-better.
        VL107 refers to VoxLingua107.
        The best and second-best results are in \textbf{bold} and \underline{underlined}.
    }
    \label{tab:hear2021}
    \begin{adjustbox}{max width=\linewidth}
    \begin{threeparttable}
    \begin{tabular}{@{}l@{}c@{~}c@{~}c@{~}c@{~}c@{~}c@{~}c@{~}c@{~}c@{~}c@{~}c@{~}c@{~}c@{~}c@{~}c@{~}c@{~}c@{~}c@{~}c@{~}c@{~}c@{~~}c@{}}
        \toprule
        & & & & & & & & \multicolumn{2}{@{~}c@{~}}{GTZAN} & & & \multicolumn{2}{@{~}c@{~}}{Mridangam} & & \multicolumn{2}{@{~}c@{~}}{\shortstack{Speech \\ Commands}} & & \multicolumn{2}{@{~}c@{~}}{\shortstack{NSynth \\ Pitch}} \\
        \cmidrule{9-10}
        \cmidrule{13-14}
        \cmidrule{16-17}
        \cmidrule{19-20}
        & Beehive & \shortstack{Beijing \\ Opera} & \shortstack{CREMA \\ -D} & \shortstack{DCASE \\ 2016$^{\spadesuit}$} & ESC-50 & FSD50K & Gunshot & Genre & \shortstack{Music \\ Speech} & \shortstack{Libri \\ Count} & \shortstack{Maestro \\ 5h$^{\spadesuit}$} & Stroke & Tonic & & 5h & full & & 5h & 50h & \shortstack{Vocal \\ Imitation} & \shortstack{VL107 \\ Top 10}  \\
        \cmidrule{2-22}
        Model & AUROC & Acc & Acc & \shortstack{Onset \\ FMS} & Acc & mAP & Acc & Acc & Acc & Acc & \shortstack{Onset \\ FMS} & Acc & Acc & & Acc & Acc & & \shortstack{Pitch \\ Acc} & \shortstack{Pitch \\ Acc} & mAP & Acc & \textbf{Avg} \\
        \midrule
        \textbf{Teacher} \\
        ~~WavLM & 61.2 & 90.2 & 76.6 & 88.7 & 73.3 & 43.7 & 93.2 & 83.4 & 96.1 & 77.2 & 12.8 & 96.7 & 89.6 & & 95.9 & 96.9 & & 34.8 & 67.4 & 14.9 & 73.8 & 71.9 \\
        ~~ATST Frame & 64.6 & \underline{95.8} & 76.7 & \textbf{95.7} & \underline{89.0} & 55.7 & 94.3 & \underline{88.3} & \textbf{100.0} & 78.1 & 24.4 & \textbf{97.5} & \textbf{96.9} & & 92.6 & 95.1 & & \underline{68.6} & 82.0 & \underline{22.3} & 66.9 & 78.1 \\
        ~~WavLM + ATST & 63.4 & 94.9 & 79.2 & \underline{95.0} & 87.8 & \underline{56.0} & 94.0 & 87.7 & \textbf{100.0} & \textbf{79.4} & 29.8 & \textbf{97.5} & 96.5 & & 96.1 & 97.1 & & 64.2 & 81.9 & 21.8 & 69.4 & 78.5 \\
        \midrule
        \textbf{Proposed} \\
        ~~USAD Small & 78.6 & 94.5 & 78.2 & 89.5 & 81.8 & 51.1 & 93.2 & 86.6 & 98.5 & 77.0 & 25.3 & 97.3 & 94.3 & & 96.2 & 97.2 & & 55.6 & 77.7 & 20.0 & 73.6 & 77.2 \\
        ~~USAD Base & 84.7 & \underline{95.8} & \textbf{80.0} & 93.6 & 82.2 & 52.2 & 94.0 & 86.3 & \textbf{100.0} & 78.7 & 26.7 & 97.3 & 95.7 & & 96.6 & \underline{97.6} & & 57.0 & 81.6 & 19.5 & \textbf{76.0} & 78.7 \\
        ~~USAD Large & \underline{86.5} & 94.1 & \underline{79.5} & 93.9 & 83.4 & 53.0 & \textbf{97.6} & 87.4 & \textbf{100.0} & \underline{79.1} & \underline{38.4} & \underline{97.4} & 96.1 & & \underline{97.0} & 97.5 & & 57.0 & \underline{83.2} & 18.5 & \underline{75.3} & \underline{79.7} \\
        \midrule
        SOTA$^{\clubsuit}$ & \textbf{87.8} & \textbf{97.5} & 75.2 & 92.5 & \textbf{96.7}& \textbf{65.5} & \underline{96.7} & \textbf{90.8} & \underline{99.2} & 78.5 & \textbf{46.9} & \textbf{97.5} & \underline{96.6} & & \textbf{97.6} & \textbf{97.8} & & \textbf{87.8} & \textbf{90.0} & \textbf{22.7} & 72.0 & \textbf{83.7} \\
        \bottomrule
    \end{tabular}
    \begin{tablenotes}[flushleft]
        \item
            $^{\spadesuit}$ Frame-level tasks.
            $^{\clubsuit}$ Each task's best result as of May 28, 2025, according to the HEAR leaderboard (\url{https://hearbenchmark.com/hear-leaderboard.html}).
    \end{tablenotes}
    \end{threeparttable}
    \end{adjustbox}
    \vspace{-4pt}
\end{table*}

%% file: table/ablation_teacher.tex
\begin{table}[t]
    \centering
    \caption{
        Phoneme error rate~(PER) and AS-20K results with USAD Small Ablation distilled from different teacher models.
        The default setting for USAD is \underline{underlined}.
    }
    \label{tab:ablation-teacher}
    \begin{tabular}{@{~}llcc@{~}}
        \toprule
        & & PR & AS-20K \\
        Speech Teacher & Audio Teacher & PER$\downarrow$ & mAP$\uparrow$ \\
        \midrule
        WavLM Base+ & -- & 7.9 & 27.7 \\
        data2vec 2.0 Speech & -- & \textbf{6.6} & 24.5 \\
        -- & SSAST Patch & 88.7 & 19.8 \\
        -- & SSAST Frame & 45.0 & 26.4 \\
        -- & BEATs iter3 & 55.5 & 33.6 \\
        -- & EAT & 58.5 & 33.8 \\
        -- & SSLAM & 60.5 & 34.2 \\
        -- & ATST Frame & 28.3 & \textbf{34.8} \\
        -- & data2vec 2.0 Audio & 25.8 & 32.8 \\
        \midrule
        WavLM Base+ & SSAST Patch & 11.0 & 28.5 \\
        WavLM Base+ & SSAST Frame & 11.9 & 29.8 \\
        WavLM Base+ & BEATs iter3 & 11.2 & \textbf{30.6} \\
        WavLM Base+ & EAT & 10.0 & 29.7 \\
        WavLM Base+ & SSLAM & 9.9 & 29.7 \\
        \underline{WavLM Base+} & \underline{ATST Frame} & 8.7 & \textbf{30.6} \\
        WavLM Base+ & data2vec 2.0 Audio & 9.5 & 30.3 \\
        data2vec 2.0 Speech & ATST Frame & \textbf{7.4} & 29.2 \\
        data2vec 2.0 Speech & data2vec 2.0 Audio & 8.2 & 29.5 \\
        \bottomrule
    \end{tabular}
    \vspace{-8pt}
\end{table}

%% file: figure/utt_ratio_per_map.tex
\begin{figure}[t]
    \centering
    \begin{subfigure}[b]{0.483\linewidth}
         \centering
         \includegraphics[width=1.03\linewidth]{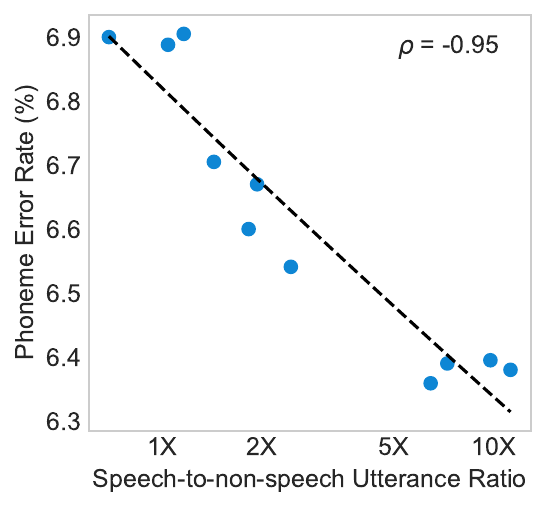}
         \vspace{-17pt}
         \caption{}
         \label{fig:utt-ratio-per}
     \end{subfigure}
     \hfill
     \begin{subfigure}[b]{0.496\linewidth}
         \centering
         \includegraphics[width=1.03\linewidth]{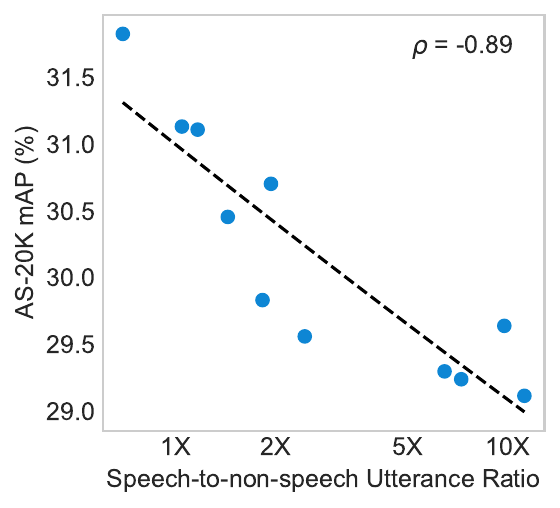}
         \vspace{-17pt}
         \caption{}
         \label{fig:utt-ratio-map}
     \end{subfigure}
    \vspace{-15pt}
    \caption{
        Training dataset speech-to-non-speech ratio vs. downstream speech~(phoneme recognition) and audio~(AS-20K) performance.
        All results are USAD Base with WavLM Base+ and data2vec 2.0 Audio teachers trained on a lower budget.
    }
    \label{fig:utt-ratio-per-map}
\end{figure}

%% file: table/distill_strategy.tex
\begin{table}[t]
    \centering
    \caption{
        Phoneme recognition and AS-20K with USAD Small Ablation and different distillation strategies.
    }
    \label{tab:distill-strategy}
    \begin{tabular}{@{~}lcc@{~}}
        \toprule
        & PR & AS-20K \\
        Method & PER$\downarrow$ & mAP$\uparrow$ \\
        \midrule
        \multicolumn{3}{l}{\textbf{Proposed}} \\
        ~~L1-cosine + no masking + $K=$ 4 Layers & 9.5 & 30.3 \\
        \midrule
        \multicolumn{3}{l}{\textbf{Distillation Objective}} \\
        ~~InfoNCE + mask-prediction~\cite{chang2024colld} & 10.4 & 29.1 \\
        ~~InfoNCE + no masking & 11.3 & 31.9 \\
        ~~L1-cosine + mask-prediction & 8.9 & 28.3 \\
        \midrule
        \multicolumn{3}{l}{\textbf{Layer-to-layer Distillation}} \\
        ~~$K=$ 12 Layers & 10.2 & 29.0 \\
        ~~$K=$ 6 Layers & 9.6 & 26.5 \\
        ~~$K=$ 3 Layers & 9.4 & 29.9 \\
        \bottomrule
    \end{tabular}
    \vspace{-8pt}
\end{table}

%% file: figure/flops_per_map.tex
\begin{figure}[t]
    \centering
    \begin{subfigure}[b]{0.49\linewidth}
         \centering
         \includegraphics[width=1.03\linewidth]{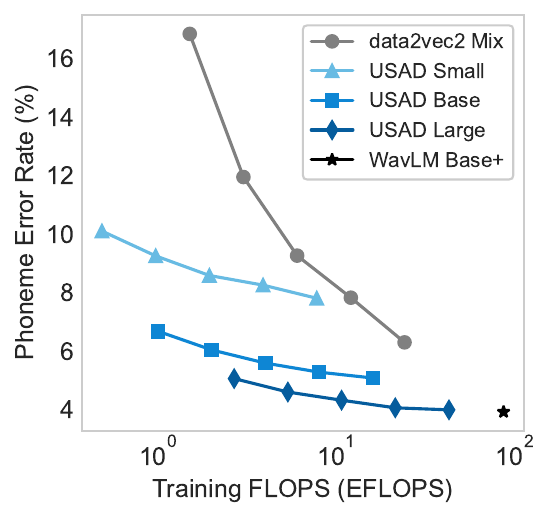}
         \vspace{-17pt}
         \caption{}
         \label{fig:flops-per}
     \end{subfigure}
     \begin{subfigure}[b]{0.483\linewidth}
         \centering
         \includegraphics[width=1.03\linewidth]{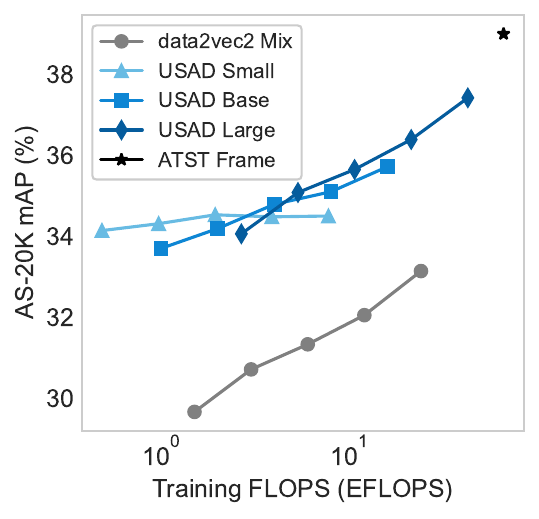}
         \vspace{-17pt}
         \caption{}
         \label{fig:flops-map}
     \end{subfigure}
     \vspace{-6pt}
    \caption{
        Training FLOPS vs. downstream speech~(phoneme recognition) and audio~(AS-20K) performance.
    }
    \label{fig:flops-per-map}
    \vspace{-4pt}
\end{figure}

%% file: section/conclusion.tex
\section{Conclusion}
\label{sec:conclusion}

We present Universal Speech and Audio Distillation~(USAD), a unified framework that learns general-purpose audio representations via distillation from domain-specific teachers.
USAD bridges speech and non-speech domains with strong performance on SUPERB, HEAR, AudioSet, and ESC-50.
Sparse layer-to-layer distillation with a frame-wise L1-cosine similarity loss enables competitive and compute-efficient models.
Future work includes improving robustness, extending to multilingual speech, and applying USAD to realistic tasks like audio large language models.